# Revealing Optical Transitions and Carrier Recombination Dynamics within the Bulk Band Structure of Bi$_2$Se$_3$

Giriraj Jnawali,[*,†] Samuel Linser,[†] Iraj Abbasian Shojaei,[†] Seyyedesadaf Pournia,[†] Howard E. Jackson,[†] Leigh M. Smith,[*,†] Ryan F. Need,[‡] and Stephen D. Wilson,[‡]

[†] *Department of Physics, University of Cincinnati, Cincinnati, OH 45221, USA*
[‡] *Materials Department, University of California, Santa Barbara, CA 93106, USA*

[*] Corresponding authors:   E-mail: giriraj.jnawali@uc.edu; Ph.: 513-556-0542 (G.J.) &
E-mail: leigh.smith@uc.edu; Ph.: 513-556-0501 (L.M.S.)

# Abstract

Bismuth selenide ($Bi_2Se_3$) is a prototypical three-dimensional topological insulator whose Dirac surface states have been extensively studied theoretically and experimentally. Surprisingly little, however, is known about the energetics and dynamics of electrons and holes within the bulk band structure of the semiconductor. We use mid-infrared femtosecond transient reflectance measurements on a single nanoflake to study the ultrafast thermalization and recombination dynamics of photoexcited electrons and holes within the extended bulk band structure over a wide energy range (0.3 – 1.2 eV). Theoretical modeling of the reflectivity spectral lineshapes at 10 K demonstrates that the electrons and holes are photoexcited within a dense and cold electron gas with a Fermi level positioned well above the bottom of the lowest conduction band. Direct optical transitions from the first and the second spin-orbit split valence bands to the Fermi level above the lowest conduction band minimum are identified. The photoexcited carriers thermalize rapidly to the lattice temperature within a couple of picoseconds due to optical phonon emission and scattering with the cold electron gas. The minority carrier holes recombine with the dense electron gas within 150 ps at 10 K and 50 ps at 300 K. Such knowledge of interaction of electrons and holes within the bulk band structure provides a foundation for understanding of how such states interact dynamically with the topologically protected Dirac surface states.

**KEYWORDS:** *$Bi_2Se_3$, mid-IR carrier dynamics, ultrafast spectroscopy, optical transition, band gap*

Three-dimensional topological insulators (3D-TIs) such as $Bi_2Se_3$ and $Bi_2Te_3$ are a new class of electronic materials which are insulating in the bulk, like ordinary semiconductors, but metallic at the surface due to Dirac surface states caused by the strong spin-orbit interaction and time reversal symmetry.[1-6] The bulk bands of these materials are occupied by massive fermions which are easily scattered by disorder or defects while the surface bands are massless fermions which are protected by time-reversal symmetry, thus providing a unique platform to be used in future electronic and spintronic devices.[7-9] Recently, topological surface states have also been recognized as having the potential to significantly enhance thermoelectric efficiency[3, 10-13] because TIs are excellent thermoelectric materials with a high Seebeck coefficient and low thermal conductivity.[14-16] The coexistence of two distinctive electronic states and strong size effects allow various tunable options in order to maximize thermoelectric efficiency in TIs.[11, 17] A key challenge is to optimize the interrelated transport and thermoelectric parameters, which requires a deep understanding of bulk electronic structure and energy relaxation processes of free carriers because topological surface states are also critically dependent on the bulk band structure of the crystal.[18-20]

Among 3D-TIs, $Bi_2Se_3$ is widely studied due to its simple surface band dispersion with a single Dirac point and robust topological character up to room temperature.[1, 5, 21] The electronic structure of $Bi_2Se_3$ has been investigated theoretically[1, 16, 22, 23] and experimentally by using several spectroscopic techniques including angle resolved photo-emission spectroscopy (ARPES).[2, 5, 22, 24-27] A unique feature in the band structure is the band inversion between valence and conduction bands due to strong spin-orbit coupling,[1, 28, 29] which opens up a gap by lifting the degeneracy at the Γ-point. The presence of edge states and time-reversal symmetry ensures the existence of chiral Dirac states between the bulk band edges; these states have been extensively mapped by ARPES.[2]

In contrast, surprisingly little is understood about the bulk band structure experimentally, in particular the photoexcited carrier dynamics. The bulk band gap measured by the ARPES technique (a surface sensitive measurement) shows a gap within a range of 0.2 – 0.3 eV by probing valence band structure and the projected bulk band edge.[2, 22, 25, 27] Discrepancies in the gap energy are attributed to the complexity with the precise location of valence band maximum at Γ-point near the surface.[22, 23, 30] Linear optical absorption spectroscopy has been used in order to characterize the bulk electronic states where estimates of the band gap vary between 0.2 – 0.35 eV.[31-35] For bulk samples the significant inconsistency of band edge parameters has been attributed to the variation in doping during sample preparation which shifts the Fermi level quite considerably.[36] One should also expect optical transitions between many higher lying valence and conduction bands. Photoexcited electrons and holes should interact dynamically with both the extended bulk band structure and also the conducting Dirac surface states. Understanding these interactions and dynamics are critical to the development of optoelectronic and spintronic devices made out of these topological materials.

Virtually all studies of carrier dynamics in $Bi_2Se_3$ have used probes of surface state carriers using time-resolved ARPES or second harmonic generation at the surface of a $Bi_2Se_3$ crystal.[37-41] Additional measurements have been made well above the band edge in relatively thin $Bi_2Se_3$ films[42-46] as well as probing the low-energy Drude response.[47, 48] In all of these studies, it has been assumed that the dominant carrier relaxation channel is through surface state bands resulting in ultrafast dynamics on the order of picoseconds.

In this work, we investigate the electronic structure and carrier dynamics in exfoliated bulk-like thick (200 – 500 nm) $Bi_2Se_3$ nanoflakes by using transient reflectance (TR) spectroscopy at 10 K and 300 K. We probe the transient response of the sample over a wide spectral range (0.3 – 1.2 eV), which captures not only the dynamic changes of interband optical transitions but also the ultrafast energy relaxation processes of hot photoexcited carriers. The measured TR spectra of these optical transitions are modeled using a simple band-filling model. The model allows us to extract the band edge parameters of our samples. Energy relaxation dynamics of photoexcited carriers are measured for different electronic states by varying the energy of the probing laser pulse. Combining our experimental results and model calculations, we discuss our findings of direct optical transitions and relaxation pathways of photoexcited carriers around bulk band edge regions of $Bi_2Se_3$.

Degenerately doped (doping density, $N_D \sim 10^{19}$ cm$^{-3}$) $n$–type single crystals of $Bi_2Se_3$ were grown from stoichiometric elements in a vertical Bridgeman furnace (see Methods).[49] Large and relatively thick nanoflakes exfoliated from a single crystal of $Bi_2Se_3$ were transferred to a high purity undoped silicon (Si) substrates with a 300 nm thermal oxide layer. Figure 1a shows a representative atomic force microscope (AFM) topography of the flake chosen for our optical investigations with height profiles and enlarged scans around the regions of interest. The regions of interest are indicated by dashed rectangles 1 and 2 each with an area of $5 \times 5$ μm$^2$. The thicknesses in these regions, as estimated from the AFM height profiles, are 250 nm and 280 nm, respectively. The surface morphology, as shown by zoomed in scans on the right, display an atomically smooth surface, without any defects or large protrusions, suggesting no noticeable surface degradation of the flake.

The samples were characterized by acquiring Raman spectra within these regions as well as from other flakes using a 632.8 nm continuous wave laser source (see Figure 1b). Three well-resolved spectrally sharp Raman peaks at $70 \pm 1$ cm$^{-1}$, $130 \pm 1$ cm$^{-1}$ and $173 \pm 1$ cm$^{-1}$ confirm the expected $A_{1g}^1$, $E_g^1$ and $A_{1g}^2$ vibrational modes, respectively.[50, 51] The absence of additional Raman features within the frequency range of $50-400$ cm$^{-1}$ supports the AFM results of clean and oxide free samples.[52]

Energy-dependent carrier dynamics are measured using the change of the reflectance of a variable mid-infrared probe pulse ($0.3-1.2$ eV) from a single $Bi_2Se_3$ nanoflake caused by photoexcitation by a 1.51 eV pump pulse (pulse width ~ 200 fs) in a typical pump-probe approach. The transient reflectance (TR) signals collected by varying the delay time $t$ between pump and probe pulses at each probe energy $E$ are normalized by the initial absolute reflectivity $R_0$ recorded before exciting the samples and denoted by $\Delta R(E,t)/R_0(E) = \{R(E,t) - R_0(E)\}/R_0(E)$. The TR signal from the oxidized Si substrate was confirmed to be negligible over the same probe energy range. Figure 2a shows 2−dimensional false color plots at both 10 K and 300 K of normalized TR signal $\Delta R/R_0$ over a wide spectral range (0.3 − 1.2 eV) and at long delay times up to 50 ps. The low-energy region (0.3 − 0.72 eV) has an order of magnitude larger signal than the high-energy region (0.8 − 1.2 eV), and so the data is reduced by $1/10$ to enable plotting both regions on the same scale. Data within the energy range of 0.72 − 0.8 eV are not available due to technical limitations. The TR maps shown in Figure 2a exhibit derivative-like spectral features and a weak modulation of signal at different delays. Two long lasting features are observed at both low (~ 0.35 eV) and high (~ 1.15 eV) energy regions at 10 K where the TR signal is strongly modulated, while the higher energy feature is substantially weaker at 300 K. The observed derivative-like features are consistent with direct interband optical transitions in the band structure. Over a broad spectral region outside these energies, the response decreases substantially. The maximum change of reflectance has magnitudes on the order of a few percent in the low-energy (~ 0.35 eV) region; the change is smaller by one order of magnitude in the high-energy (~ 1.15 eV) region with identical excitation fluencies. In addition to the derivative-like spectral features, a strong oscillatory modulation of the signal as a function of delay is also apparent in the higher energy region which corresponds to coherent longitudinal acoustic (LA) phonons in $Bi_2Se_3$ crystals.[42, 43, 53]

A clearer view of these major spectral features can be observed by plotting horizontal spectral slices of the transient data at various delay times. A number of such slices at different delay times for the 10 K and 300 K data are shown in Figure 2b. Except for a slight decrease in magnitude, clear derivative-like features at the low- and high-energy regions consistently appear for all delay times, providing directly the time evolution of interband optical transitions in the electronic band structure of bulk $Bi_2Se_3$. The derivative-like feature at low energies shows a typical decrease in reflectivity on the low-energy side of the lineshape. In contrast, the derivative-like feature in the higher energy near-IR region is phase-shifted by 180°, showing a strong positive increase of the TR signal on the low-energy side of the spectral lineshape. We note that both 10 K and 300 K spectra show nearly identical lineshapes and spectral positions of the low-energy feature. No broadening of the 300 K lineshape is observed, and virtually no shift with temperature. Additionally, we note that the high-energy feature is hardly visible at room temperature because of the fast relaxation of the high-energy carriers in contrast to the longer relaxation times at low temperatures.

Figure 2c displays a detailed look at the low-energy spectral lines over the first 5 picoseconds. These spectra are distorted spectrally and blue-shifted immediately at zero delay but recover back towards lower-energies within a couple of ps. At times following 5 ps after the pump pulse the spectral lineshapes remain unchanged afterwards for several hundreds of picoseconds. These ultrafast energy shifts of the absorption minima as a function of time at both 10 K and 300 K are shown in Figure 2d. Compared with the stable lineshape at later times, the peak of the zero time lineshape for the low-energy peak is shifted by 20 meV at 10 K and 10 meV at 300 K. By comparison, the high-energy near-IR zero time lineshape at 10 K shifts by 10 meV at zero time. At both temperatures and for both high- and low-energy features, the lineshape energy positions relax linearly with time back to the stable form within < 5 ps.

As noted previously, a number of measurements which probe the topologically protected Dirac surface states using time-resolved ARPES,[37, 38] frequency doubling,[40] sub-gap mid-IR reflectivity[54] and THz scattering[47, 48] have been reported. All of these measurements used a 1.5 eV pump since it has been shown to interact directly with the second Dirac surface state at higher energy.[45, 55] In the experiments described here, while the 1.5 eV pump pulse does interact with the second surface state, it also directly injects electrons and holes into high-energy states in $VB_1$ and $CB_1$, and also into $VB_2$ and $CB_2$ as labeled by the red arrows in Figure 3a. The lower energy tunable probe pulse is only interacting with the bulk band states since it is at a much higher energy than the first surface state and much lower energy than the second surface state. However, earlier investigations of thickness dependent photoinduced charge transport have confirmed bulk carriers completely overwhelm surface states in thicker $Bi_2Se_3$ films, *i.e.*, thicker than 50 nm.[48] Because all $Bi_2Se_3$ flakes studied are thick ($d > 200$ nm), we are confident that the majority of the dynamic response measured here at all probe energies are from perturbations of the bulk band states and not the conducting surface states.

The excitation pump pulse is focused on an area ~ 2 μm in diameter on the nanoflake. The nanoflake is expected to be heavily $n$–type because of Se vacancies, causing Fermi level shift well above the conduction band minimum $CB_1$. We estimate that the pump pulse excites approximately $10^{17}$ cm$^{-3}$ hot electron-hole pairs high up into the conduction and valence bands while the $CB_1$ already contains a high density electron gas ($N_D \sim 10^{19}$ cm$^{-3}$).[49] The photoexcited electrons are thus only a small perturbation to the dense electron gas. As illustrated in Figure 3a, the photoexcited hot electrons and holes relax towards the band minima through a rapid cascade of optical phonon emissions within a picosecond, resulting in quasi-thermalized hot electrons near the electron gas at the Fermi level in $CB_1$ and hot holes near the maxima of $VB_1$. Carrier-carrier scattering at the band minima causes rapid equilibration of the photoexcited electrons and holes with the dense electron gas. This likely is responsible for the complex distortions of spectral lineshape at the earliest times ($t < 5$ ps), as shown in Figure 2c. The subsequent steady decrease of spectral amplitude as a function of delay is caused by band-to-band carrier recombination, which will be discussed later.

The TR lineshapes result from changes in the real and imaginary part of the index of refraction caused by the presence of the photoexcited electrons and holes. The additional electrons cause a small increase in the Fermi energy in $CB_1$, which changes the energy onset of the absorption for optical transitions from the top of $VB_1$ or $VB_2$ to the Fermi level in $CB_1$. We calculate this change in the absorption coefficient (see Methods Section) which causes a corresponding change in the real part of the index of refraction calculated using a Kramers-Kronig transformation. The free parameters in this calculation are: (1) the density of the doped electron gas in $CB_1$, (2) the density of photoexcited carriers (electron-hole pairs), (3) the temperature of the electrons and holes, and (4) the fundamental gap energy $E_{g1}$ ($VB_1 \rightarrow CB_1$) as well as second (higher energy) gap $E_{g2}$ ($VB_2 \rightarrow CB_1$). We consider parabolic bands with electron and hole effective masses, $0.12\, m_e$ and $0.24\, m_e$, respectively ($m_e$: rest electron mass) for our model calculations.[32, 33] Figure 3b,c show experimental TR spectra around low- and high-energy regions which are well reproduced by the theoretical lineshapes. The dashed line shows the change in absorption coefficients due to the pump laser excitation. The low-energy lineshape shows a decrease in the absorption (photoinduced bleaching), while the high-energy lineshape indicates an increase in the absorption (photoinduced absorption).

By fitting the lower and higher energy lineshapes at 10 K, we show that the doped electron gas density is $7.5 \times 10^{18}$ cm$^{-3}$, which implies a Fermi energy of $E_f \sim 116$ meV and a fundamental $VB_1 \rightarrow CB_1$ gap of $E_{g1} \sim 0.18$ meV and a higher energy $VB_2 \rightarrow CB_1$ gap of $E_{g2} \sim 0.98$ meV. The photoexcited carrier density at the fundamental band edge is shown to be $\Delta N \sim 10^{17}$ cm$^{-3}$, which is $\sim 1\%$ of the doped electron density at a delay time of 20 ps. These parameters result in zero crossing of the TR lineshape at 0.35 eV for $E_1$ and at 1.15 eV for $E_2$ transitions. The estimated low- and high-energy gaps also can be easily observed in the minima and maxima of the $\Delta\alpha$ peaks shown in Figure 3b,c and are consistent with recent steady-state measurements on bulk samples.[31-33, 56]

While we can successfully calculate the TR lineshapes when the photoexcited electrons and holes are in thermal equilibrium with the doped electron gas, we have been unable to fit the initial relaxation of hot carriers, which occurs in the first 0 – 5 ps. As described previously, we excite electrons and holes with an energy of 1.5 eV, which is 1150 meV higher energy than the optical gap measured from optical transitions from the VB to the Fermi level in the conduction band (CB). These hot carriers relax close to the band edges (VB minimum or Fermi level in the CB) through emission of a cascade of optic phonons within ~1 ps until their energy is within a single optical phonon (~ 20 meV, see Figure 1) of the band edges.[37, 38] Most of this excess energy goes into the lattice, but there is not enough energy to raise measurably the lattice temperature since the flake is in good thermal contact with the substrate. The TR lineshapes from 0 – 5 ps thus are measuring the hot carriers after their relaxation through optic phonon emission, but before they relax to the lattice temperature. TR lineshape analysis shows that the density of these hot carriers is only $10^{17}$ cm$^{-3}$, which is only 1% of the cold degenerate Fermi electron gas (~$10^{19}$ cm$^{-3}$). Our assumption is that the hot carriers efficiently thermalize within this 5 ps through carrier-carrier scattering with the cold electrons near the Fermi level, but this transfer of energy negligibly impacts the temperature of the cold electron gas.

The early time (t < 5 ps) results shown in Figure 2c,d are consistent with the description above. The lineshape at earliest time is 20 meV above the later lineshapes, consistent with a carrier temperature no higher than 200 K above the lattice temperature. This shift relaxes from 20 meV down to zero within the first 5 ps as shown in Figure 2d, which gives an estimate of the thermal relaxation time of the hot carriers. Finally, the cold electron gas is negligibly affected by this process since the lineshape is essentially unchanged apart from amplitude from 5 ps after photoexcitation until complete decay of the photoexcited carriers.

We remark parenthetically that our situation is substantially different from the measurements made recently in graphene.[57, 58] In graphene the optical phonon energies are 10 times larger (200 meV vs. 20 meV), and the electron-phonon coupling is an order of magnitude smaller than in $Bi_2Se_3$.[59, 60] These differences mean that in graphene (unlike here) the perturbation of the laser (either by direct heating by THz radiation or by photoexcitation of hot carriers) is large enough to directly heat the entire Fermi gas. Thus, the initial temperatures observed for graphene are also much larger (*e.g.*, 2000 K in graphene, *vs.* 200 K here).

While the lineshapes described previously do not change with time after 5 ps, the amplitude of the signal does decrease with time as the photoexcited carriers recombine. Figure 4a displays long delay time traces of the TR signal measured both at 10 K and 300 K at different probe energies following 1.51 eV pump excitation. The first noticeable feature in the data is the strong non-oscillatory TR signal overlaid with a weak and damped oscillatory signal. The oscillations are weakly visible in the low-energy region (0.3 – 0.7 eV) but stronger around the high-energy region (~1.0 eV), suggesting a wavelength dependent behavior. The second important feature is the variation of the decay dynamics as a function of probe energy. The TR signal at all probe energies exhibits ultrafast relaxation at early times followed by a longer decay response, suggesting two decay channels in the relaxation processes. This feature can be seen more clearly in the logarithmic plots of time decays as a function of energy in Figure 4b, where $\tau_1$ and $\tau_2$ label the fast and slow decay times, respectively. To analyze the probe energy dependent decay behavior, each TR trace was fitted after zero time delay by using bi-exponential functions combined with an exponentially decaying cosine function as expressed below:

$$\Delta R/R_0 \simeq A_1\left(e^{-t/\tau_1}\right) + A_2\left(e^{-t/\tau_2}\right) + A_{\text{ph}}\left(e^{-t/\tau_{\text{ph}}} \cdot \cos(\{2\pi f_0 \cdot t\} + \phi)\right) + B_0. \qquad (1)$$

The first and second terms with amplitudes $A_1$ and $A_2$ and their fast $\tau_1$ and slow $\tau_2$ decay times characterize the non-oscillatory electronic signal. The third term, a damped sinusoidal function with amplitude $A_{ph}$, average frequency $f_0$, initial phase $\phi$ and damping time $\tau_{ph}$ describes the oscillatory signal caused by coherent LA phonons. The last term $B_0$ is the constant attributed to the much slower ($t > 2$ ns) and much weaker background response. A least squares fit of each trace is also overlapped on each plot, which provides the relevant dynamic parameters including the fast $\tau_1$ and the slow $\tau_2$ decay times (lifetimes) and an average phonon frequency $f_0$ (or average oscillation period, $T_0$).

First, we discuss the behavior of the fast and slow decay times at different energies. Decay times are plotted versus the probe energies in Figure 4c in which blue and green colors are used to differentiate 10 K and 300 K data, respectively. At 10 K, the early fast decay $\tau_1$ is found to increase from ~5 ps at ~0.7 eV to ~10 ps around the lower $E_1$ and upper $E_2$ transition energies, *i.e.*, 0.35 and 1.15 eV, respectively. The second decay time $\tau_2$, however, substantially increases to ~150 ps at the low-energy region (~0.3 eV), and ~500 ps at the high-energy region (~1.2 eV). At 300 K, the dependence of both the decay times with probe energies is markedly different. Both the $\tau_1$ and $\tau_2$ decay times show a maximum at the lowest energies and decrease with higher probe energy. For example, the second decay time $\tau_2$ decreases from ~50 ps around the $E_1$ transition to ~30 ps around the $E_2$ transition.

The initial rapid thermal relaxation is responsible for the ultrafast $\tau_1$ time decay. The longer $\tau_2$ decay measures the gradual recombination of the photoexcited holes in $VB_1$ and $VB_2$ with the electron gas (which includes both doped and photoexcited electrons). The lowest energy $\tau_2$ lifetime of 150 ps at 10 K therefore reflects the recombination lifetime of the ground state photoexcited holes in $VB_1$ with the dense electron gas in $CB_1$. Based on a doping density of $N_D \sim 7.5 \times 10^{18}$ cm$^{-3}$, we can estimate the upper limit of the coefficient for radiative recombination (the $B-$ coefficient) to be $B = 1/(N_D \times \tau_2) \sim 8.5 \times 10^{-10}$ cm$^3$/s, which is about a factor of three larger than in gallium arsenide (GaAs) for similar doping conditions.[61, 62] The $\tau_2$ lifetime is seen to decrease to ~50 ps at 300 K and may reflect either that the B–coefficient decreases at higher temperatures[63] or that nonradiative recombination becomes more dominant at higher temperatures.

The most interesting feature in the dynamics is the surprisingly long decay times seen only at low temperatures in the higher band transition around the $E_2$ region. The TR signal at these higher energies is less than 10 % that at the band edge. The longer lifetime suggests that there is a source of either long-lived electrons or holes which are at higher energy. The relevant energies can be understood by the bulk band structure of $Bi_2Se_3$, as schematically presented in Figure 3a. As described earlier, three allowed interband transitions are involved within the measured probe energies, as symbolized by $E_1$, $E_2$, and $E_3$. While the lower energy TR responses can only result from optical transitions from $VB_1$ to the Fermi level in $CB_1$ (shown as $E_1$ in Figure 3a), the higher energy TR measurements potentially can sample either $VB_2$ to the Fermi level in $CB_1$ ($E_2$–transition) or potentially promoting electrons from $CB_1$ to $CB_2$ ($E_3$–transition). Therefore, the longer-lived decay seen at low temperatures either reflects long-lived holes at energies near $VB_2$ or long-lived electrons near $CB_2$. Since carrier scattering from higher to lower energetic states within conduction bands has been reported to occur within a few ps at 70 K,[37] we conclude that long-lived holes at $VB_2$ contribute to the long-lived decay response at higher probe energies at low temperatures. Nevertheless, the fact that this long-lived decay response is not seen at 300 K suggests that either a small thermal barrier suppresses the relaxation of these electrons or holes, or the larger thermal population of phonons provides an efficient relaxation mechanism through scattering.

The behavior of the coherent LA phonons is reflected by the oscillating time decays as a function of probe energy. Oscillations in optical reflectance are due to interference of the probe beam reflected from the surface or interface and the strain pulse produced by the pump which propagates at the acoustic sound velocity.[64] As expected (see Figure S1 in Supporting Information), the period of the oscillations $T_0$ increases linearly with the probe wavelength. We estimate (see details in Supplementary Note 2) the average speed of the LA phonon to be $v_s \sim 2{,}500$ m/s, which is consistent with other measurements for thick or bulk samples.[53] The oscillations are pronounced around 1.0 eV because the period of the oscillations is smaller than the slower decay time $\tau_2$. They are more difficult to see near 0.35 eV when the oscillation period approaches the lifetime of the photoexcited carriers.

The central result of these experiments is that the optical transitions in $n$–type degenerately doped $Bi_2Se_3$ are determined by transitions between $VB_1$ and $VB_2$ and the Fermi level within $CB_1$. We find that the fundamental $VB_1 \rightarrow CB_1$ gap of $E_{g1} \sim 0.18$ eV combines with the degenerate $CB_1$ Fermi energy of $E_f \sim 0.116$ eV to present an optical gap of 0.356 eV. This is consistent with a doped electron concentration of $7.5 \times 10^{18}$ cm$^{-3}$. By similarly analyzing the higher energy optical transition we determine that the splitting between $VB_2$ and $VB_1$ (*i.e.*, energy difference between $E_{g2}$ and $E_{g1}$) is 0.8 eV. These parameters for the energies of the band edges are consistent with various ground-state optical spectroscopy results[31, 32, 35, 56] and several LDA/GW band structure calculations.[22, 23]

We note that the optical band gap observed in our samples is nearly twice the fundamental energy gap of 0.18 eV, which is caused by a doping-induced Moss-Burstein shift due to band-filling effects.[65-67] Such doping-induced shifts have been also reported previously in different $Bi_2Se_3$ samples.[31, 34-36] As noted earlier, unintentional high doping density with $n$–type carriers is well known in $Bi_2Se_3$ samples due to Se vacancies.[68-70] Moreover, the current-voltage (*I-V*) characteristics measured in the devices made with identical $Bi_2Se_3$ flakes (see Figure S3a in Supporting Information) display a linear *I-V* curve, suggesting a metallic character arising from the high doping density in our samples. The estimated resistivity $\sim 18$ $\Omega\mu$m with the typical (lower-end) carrier mobility of similarly synthesized samples[71, 72] $\mu_e \sim 500$ cm$^2$/V.s provides a rough estimation of carrier density, $N \sim 7 \times 10^{18}$ cm$^{-3}$, which agrees well with the doping density $N_D \sim 7.5 \times 10^{18}$ cm$^{-3}$ and the Fermi energy $E_f = 0.116$ eV estimated by model fitting of TR spectra. In addition the device also exhibits photocurrent onset at ~ 0.35 eV, which is weakly temperature dependent (see Figure S3b), qualitatively supporting the optical gap obtained by TRS data both at 10 K and 300 K (see also Supplementary Note 3).

Our data show that the thermalization and recombination dynamics are strongly impacted by the degenerate doping of the $Bi_2Se_3$. After photoexcitation by the pump pulse, hot electrons and holes come to equilibrium with the cold degenerate electron gas within ~ 2 ps because of ultrafast carrier-carrier scattering. After reaching thermal equilibrium with the lattice, the minority carrier holes recombine within 150 ps at low temperatures. This provides an upper bound to the radiative recombination, *i.e.*, B – coefficient, which is at least a factor of three larger in $Bi_2Se_3$ than for GaAs.[61, 62] It is interesting to note that the recombination time for the higher energy transition $VB_2 \rightarrow CB_1$ is substantially longer at low temperatures probably reflecting long-lived holes near $VB_2$. At room temperature these higher energy states decay much more rapidly than the band edge states. The complexity of the higher lying transition is also reflected by the fact that it results from a photoinduced absorption (increase in the absorption) in contrast to the photoinduced bleaching observed near the optical gap.

These results provide a baseline understanding of the behavior of electrons and holes in thick degenerately doped $Bi_2Se_3$. With this understanding, it may be possible to understand the coupling of the topologically protected surface states with the bulk states described here when the surface states can no longer be neglected, *e.g.*, for thinner materials. Additionally, it would be useful to see how the dynamics is changed for non-degenerately doped material, which should strongly affect the minority carrier recombination lifetime as well as thermalization times. This new understanding provides a basis for optimizing topological insulators for new physics and new technologies where one can control both the surface states and also their interactions with the bulk states.

## Methods

**Growth and exfoliation of samples:** Single crystals of $Bi_2Se_3$ were grown using the Bridgman method starting from a stoichiometric ratio of elemental bismuth (Bi) and selenium (Se) powders. The powders were ground together in an inert argon atmosphere then sealed in a quartz ampoule under $10^{-5}$ mbar vacuum. Growth began with a 24 hours hold at 840°C to ensure complete melting and homogeneity of the precursor powders. The ampoule was then translated at 2 mm/hr over 5 d through an average zone gradient of 10 C/cm. Final sample annealing was done holding the solidified material at 500°C for 24 hrs. Thin flakes of $Bi_2Se_3$ samples were prepared by mechanical exfoliation of these $Bi_2Se_3$ crystals on an oxidized and low-doped silicon substrates (300 nm $SiO_2$/Si) using a "Scotch tape" method similar to that used for graphene.[73] The sample was then quickly mounted onto the cold finger of a liquid helium cooled cryostat (Janis ST-500) for optical measurements, which operates between 8 K and 300 K by maintaining the sample under high vacuum conditions (pressure, $p < 1 \times 10^{-4}$ Torr).

**Sample characterization**. Surface morphology and thickness of $Bi_2Se_3$ nanoflakes were measured by using atomic force microscope (AFM). Measurements were conducted in air on a Veeco AFM using gold (Au) coated platinum (Pt) cantilevers operating in tapping mode. Raman spectroscopy was applied to characterize film quality and possible degradation of sample by oxidation. The measurements were performed using a Dilor triple grating spectrometer coupled to a microscope with 632.8 nm laser excitation in the backscattering configuration under ambient conditions. A $100\times$ objective was used to focus the incident beam and collect the scattered signal and then dispersed by 1800 g/mm grating after the resonant scattered laser light is removed using a double subtractive mode spectrometer. The Raman scattered signal is measured with a spectral resolution of 1 cm$^{-1}$. The lowest frequency observable is 50

cm$^{-1}$. Laser power was maintained around 100 μW during the measurements to minimize laser-heating effect in our samples.

**Transient reflectance spectroscopy**. Femtosecond pump-probe TR measurements were performed based on a Ti:sapphire oscillator which produces 200 fs pump pulses at a central wavelength of 800 nm with 80 MHz repetition rate and an average power of 4 W. The majority (80 %) of the laser beam was used to pump an optical parametric oscillator (OPO), which generates variable wavelength signal beam (0.8 − 1.2 eV) and an idler beam (0.3 − 0.72 eV). All measurements were performed using photoexcitation at 820 nm with a photon energy of $\hbar\omega_p = 1.51$ eV and probing with both signal and idler outputs, respectively. The probe pulses are delayed in time with respect to the pump pulses by using a motorized linear translation stage. The pump and probe beams are spatially overlapped on a single flake using a 50× protected silver reflective objective with the help of a CCD camera, and the reflected beams are directed to a liquid nitrogen cooled InSb detector. The pump beam is filtered out during the TR measurements using a long pass filter. The signal is collected with a lock-in amplifier phase-locked to an optical chopper that modulates the pump beam at a frequency of 1 kHz. The focused spot size of both pump and probe beams are a nearly identical ∼ 2 μm in diameter. An average pump fluence of ∼ 1 mJ/cm$^2$, which is still well below the damage threshold in bulk Bi$_2$Se$_3$, was used throughout the experiments, while the probe fluence was kept nearly one order of magnitude lower, *i.e.*, ∼ 0.1 mJ/cm$^2$.

**Modeling transient reflectance spectroscopy**. Our model is primarily based on photoinduced state filling contributions to the change of the dielectric response: a fractional change in the complex refractive index ($\tilde{n} = n + ik$), which is measured in the form of change in reflectance signal at different probe energies. The imaginary component of the complex refractive index relates to the absorption coefficient as: $k = (\lambda/4\pi)\alpha$. Hence, we begin the model by calculating the photoinduced change in the absorption.

The carrier dependent absorption coefficient between parabolic bands can be described using the following expression:[74, 75]

$$\alpha(E, N, T) = \frac{C_\alpha}{E} \int_0^{E-Eg} \rho_c(E') \\ \cdot \rho_v(E' - E) \times [f_l(E - E_g - E', N, T) - f_u(E', N, T)] \, dE' \qquad (2)$$

where $N$ and $T$ denote carrier density and their temperature, $\rho_c$ and $\rho_v$ are density of states for the conduction and valence bands, $f_l$ and $f_u$ are the Fermi-Dirac distribution functions for carriers occupying upper and lower electronic states and $C_\alpha$ is a constant factor fit to the data. We assume the photoexcited electrons amount to only a small (~1%) increase in total electron density, since the samples are heavily n-doped. The photoexcited hole density is taken to be equal to the photoexcited electron density. Since the photoexcited carriers thermalize within a couple of picoseconds timescale to their respective band minima and then scatter rapidly with the dense and cold electron gas, we take the carrier temperature to be equal to the lattice temperature for the fits presented in this work.

Once the modulation in the absorption, $\Delta\alpha$, is calculated, we can derive the modulation of the real part of index of refraction, $\Delta n$, using the Kramers-Kronig relation:

$$\Delta n(E, N, T) = \frac{\hbar c}{\pi} \int_0^\infty \frac{\Delta\alpha(E', N, T)}{(E')^2 - E^2} dE' \qquad (3)$$

where $\Delta\alpha$ is calculated from Eq. 2. Note that although the upper bound of the integral is infinite, $\Delta\alpha$ rapidly approaches zero above the band edge and thus our theoretical description of the absorption need not include the high-energy regime.

Assuming normal incidence of the probe beam, the reflectance of the sample without excitation is given by:

$$R_0 = \frac{(n-1)^2 + k^2}{(n-1)^2 - k^2}. \tag{4}$$

The fractional change of reflectance induced by pump excitation (normalized by initial reflectance $R_0$ before exciting the sample) can be expressed in first-order expansion as following:

$$\frac{\Delta R}{R_0} \cong \left(\frac{8 n_o k_o}{((n_o + 1)^2 + k_o^2)^2}\right) \cdot \Delta k(E, N, T) - \left(\frac{4(1 + k_o^2 - n_o^2)}{((n_o + 1)^2 + k_o^2)^2}\right) \cdot \Delta n(E, N, T) \tag{5}$$

where $n_0$ and $k_0$ are the background values of the real and imaginary components of the complex index of refraction, respectively. We take $n_0$ to be a constant at the average value of 5.5, based on ellipsometry measurements on bulk $Bi_2Se_3$.[76] We calculate $k_0 = (\lambda/4\pi) \cdot \alpha$ from Eq. 2 using background estimates for carrier density.

We fit theoretical lineshapes from Eq. 5 to our experimental spectra by parameterizing the density of doped electrons, density of photoexcited carriers, band gap energy, $C_\alpha$, and an overall scaling factor. We numerically optimize these parameters to minimize the sum squared error for spectra taken at distinct delay times, keeping the time-independent parameters consistent. We fit the low-energy spectra first to establish the doping density, and then fit the high-energy spectra with this value fixed. The resulting parameters and their associated uncertainties are tabulated in the Supplementary Note 1.

## Acknowledgments

We acknowledge the financial support of the NSF through grants DMR 1507844, DMR 1531373 and ECCS 1509706. S.D.W. acknowledges the support of NSF DMR 1505549.

## Author contributions

G.J. and L.M.S. conceived the study. R.F.N. and S.D.W. grew single crystals of $Bi_2Se_3$. G.J. prepared and characterized nanoflake samples, performed TRS measurements, data analysis, figure planning, and draft preparation. S.L. and I.A.S. helped in the TRS experiments. S.L. modeled the TRS spectra with inputs from L.M.S., G.J. and H.E.J. S.P. fabricated single flake devices and performed photocurrent measurements under the guidance of G.J., L.M.S. and H.E.J. G.J. and L.M.S. wrote the manuscript with inputs and comments from H.E.J. and S.L. All coauthors contributed to the discussion of results and commented on the manuscript.

## Supporting Information Available

Ground- and excited-state band parameters, analysis of coherent longitudinal acoustic (LA) phonon, modeling TRS spectra at 300 K and photocurrent spectroscopy results of a $Bi_2Se_3$ device are described in Supplementary Notes 1–3. Wavelength dependent LA phonon period, model fittings of TRS spectra at 300 K, *I-V* curves and photocurrent spectra of a single $Bi_2Se_3$ nanoflake device at 10 K and 300 K are presented in Figures S1–S3.

## Corresponding authors:

*E-mail: giriraj.jnawali@uc.edu (G.J.)
*E-mail: leigh.smith@uc.edu (L.M.S.)

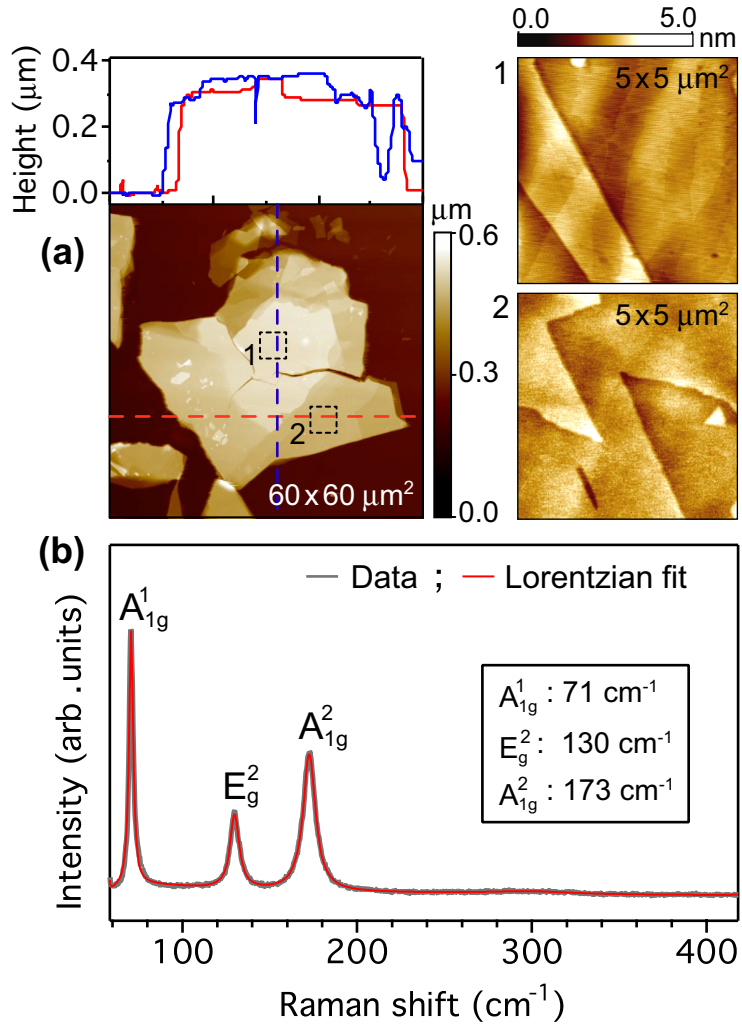

**Figure 1.** Morphology and structure of a thick $Bi_2Se_3$ nanoflake. (a) AFM topography and surface morphology characterization of the representative $Bi_2Se_3$ flake that has been used for optical measurements. The height profiles were taken along horizontal and vertical directions on the flake as shown by dashed lines with respective colors (red and blue lines, respectively) in the AFM image. Nominal thickness around the regions marked by dashed rectangles 1 and 2, where most of the optical measurements were performed, is ~ 250 nm and ~ 350 nm, respectively. Surface morphology on the regions 1 and 2 are shown in the zoomed in images (increased contrast) in the right panel, which exhibits atomically flat surface (from single layer to few layers high steps) without any noticeable large protrusions, suggesting atomically clean surfaces. (b) Representative Raman spectrum around these regions showing the three expected Raman active optical phonon modes associated with in-plane and out-of-plane lattice vibrations.

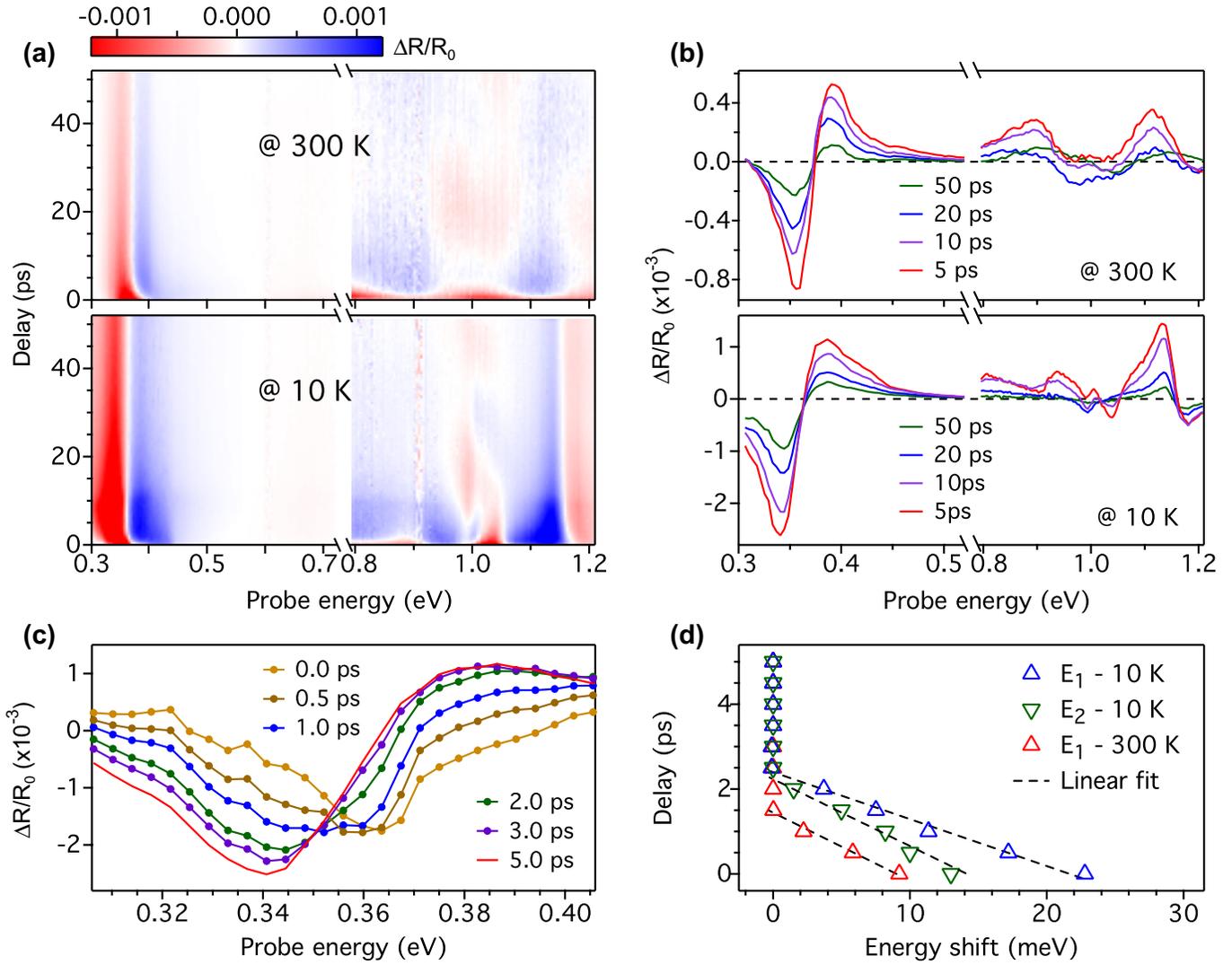

**Figure 2.** Transient mid-IR reflectance response of a photoexcited $Bi_2Se_3$ nanoflake. (a) Two dimensional false color plots of transient reflectance (TR) signal ($\Delta R/R_0$) measured at 10 K (lower panel) and 300 K (upper panel). A pump laser pulse of photon energy of 1.51 eV was used to excite the sample and the induced dynamic change of optical response is measured by collecting the reflected signal of a variable energy probe pulse as a function of delay relative to the pump pulse at each probe energy. (b) Spectral slices extracted from horizontal line cuts at different delay times in corresponding plots of (a). The spectra show two strong and long-lived features at around the 0.34 eV and 1.2 eV, corresponding to fundamental optical absorption edge and higher energy band-edge transitions, respectively. In addition to spectral features of interband transitions, a periodic feature appears in time domain around the higher energy region, which is caused by longitudinal coherent acoustic phonons in the sample. The TR data in (a) and (b) in the long wavelength region between $0.3 - 0.72$ eV are reduced by one-tenth of its measured value to enhance the visibility contrast across the entire probe region. (c) The TR spectra below 5 ps delay time at 10 K, showing a dynamic blue shift of band-edge spectra. (d) Extracted energy shift as a function of delay time, which shows linear decrease of band-edge transition energy towards equilibrium energy after a dynamic Moss-Burnstein shift due to band filling. The largest shift of $E_1$ is $\sim 20\pm 2$ meV at time zero and relaxes within $\sim 2$ ps at 10 K while it is only $\sim 10$ meV at 300 K.

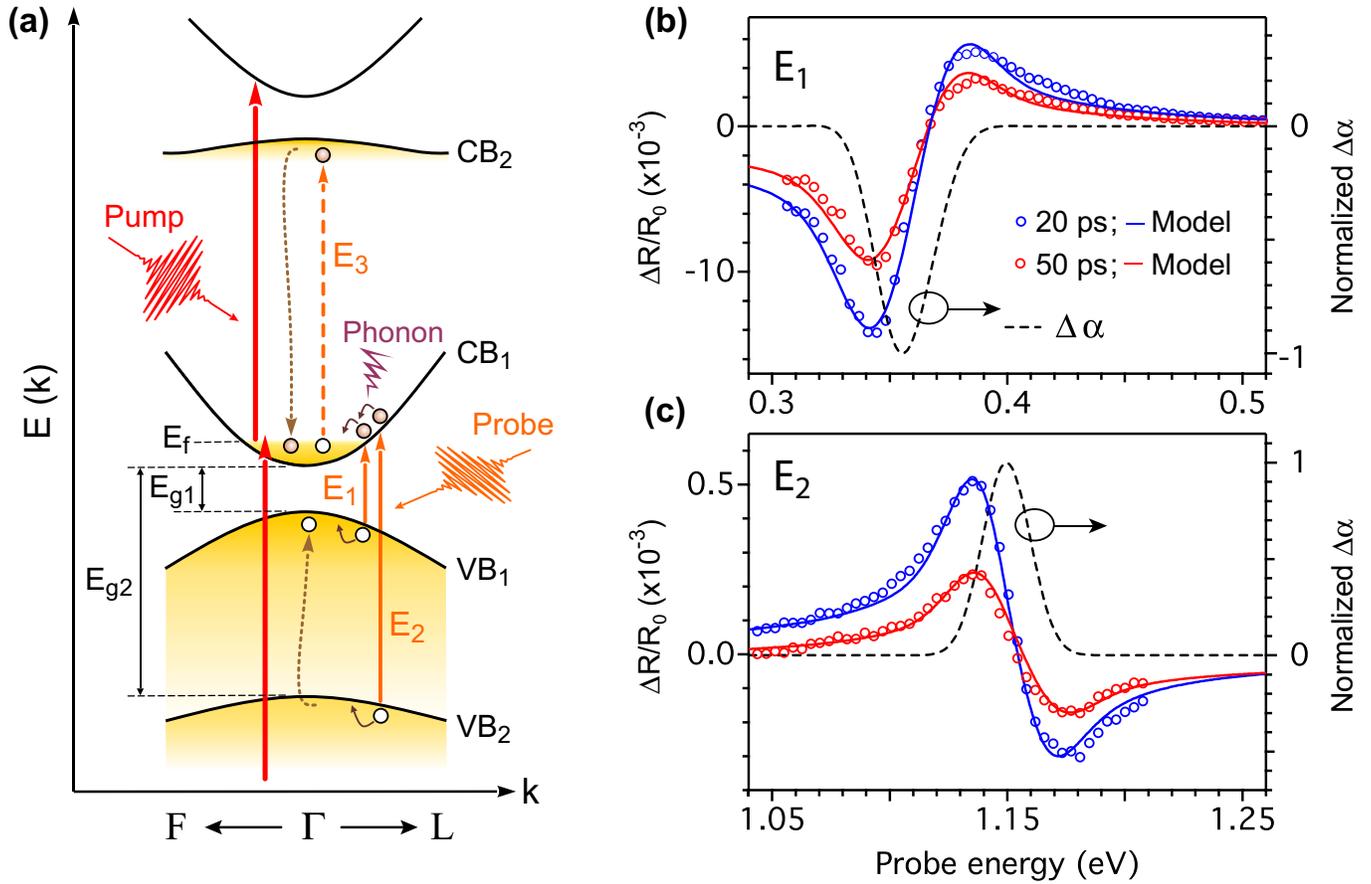

**Figure 3.** Mid-IR photoinduced processes within bulk band structure of $Bi_2Se_3$. (a) Schematic view of possible direct optical transitions of ultrafast thermalized carriers at different energetic states around the zero momentum (Γ- point) of bulk band structure of $Bi_2Se_3$ crystal. The red arrow indicates the initial photoexcitation by a 1.51 eV pump pulse. Brown dotted arrows show the ultrafast interband decay of hot carriers (electrons in the CB and holes in the VB). Carriers are rapidly thermalized to band minima by emitting optical phonons (intraband cooling). Vertical pointed orange arrows show probe-induced interband transitions near the band edge $E_1$ ($VB_1 \rightarrow CB_1$) and higher lying second band $E_2$ ($VB_2 \rightarrow CB_1$). Fundamental band-edge and higher energy gap (gap between maximum of $VB_2$ and minimum of $CB_1$) are symbolized by $E_{g1}$ and $E_{g2}$, respectively. A likely higher energy inter-CB transition is also indicated by $E_3$ ($CB_1 \rightarrow CB_2$). Model fittings (solid lines of respective colors) of band edge feature, $E_1$ (b) and high-energy feature, $E_2$ (c) of TR spectra measured at 20 (blue circle) and 50 ps (red circle) delay times at 10 K. The calculated change in absorption Δα at 20 ps normalized by the peak value is also superimposed with scale on the right (black dashed line).

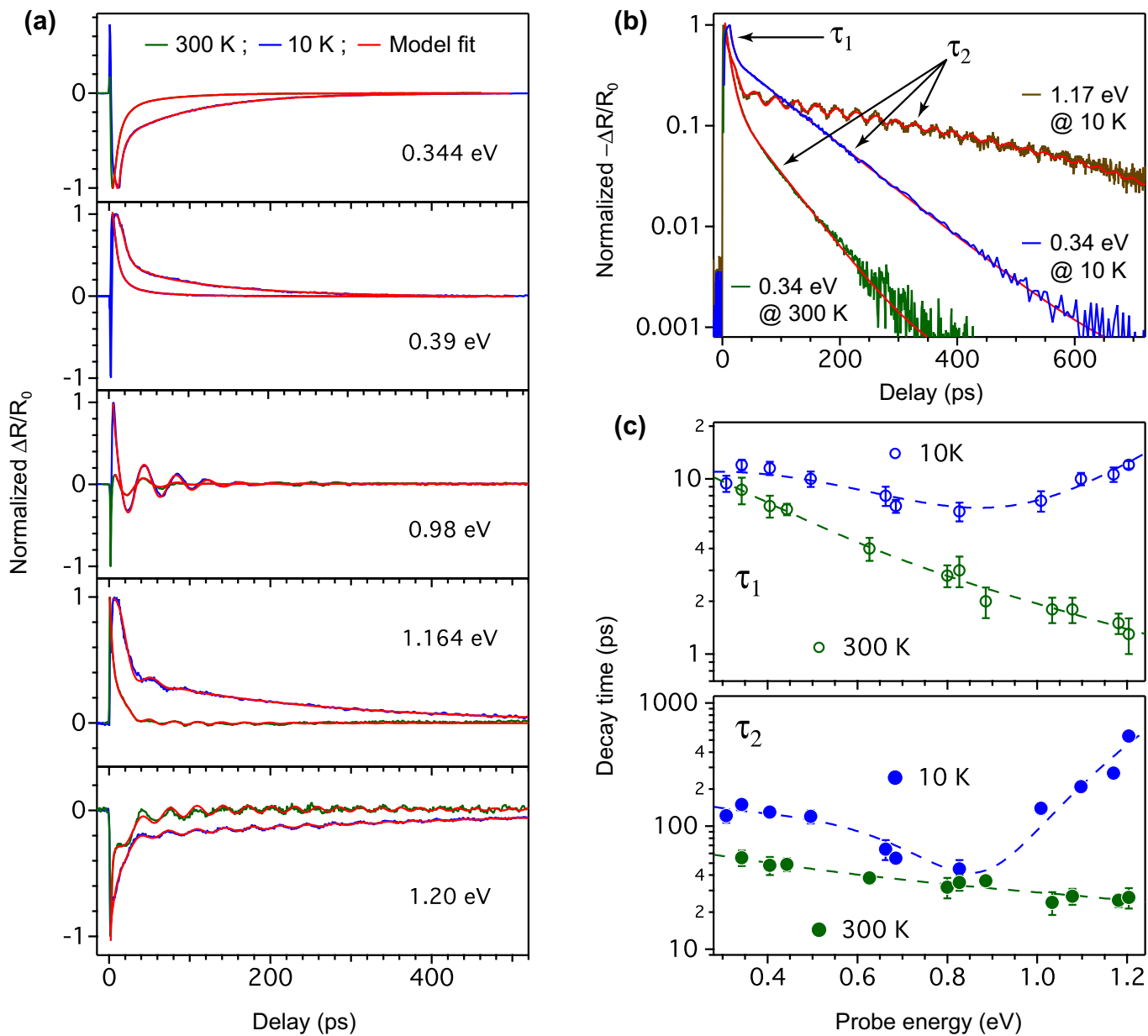

**Figure 4.** Decay dynamics of photoexcited carriers at different bulk electronic states of $Bi_2Se_3$. (a) Time traces of transient reflectance response $\Delta R/R_0$ of a $Bi_2Se_3$ flake measured at different probe energies at 10 K (blue line) and 300 K (green line). Each trace is normalized by the peak value around zero delay to compare their decay responses. The TR signal is embedded with a weak oscillatory signal due to the coherent acoustic modes generated in the sample. Each trace is fitted with a model of bi-exponential functions combined with exponentially decaying cosine function to extract decay parameters. (b) Logarithmic plots of the TR traces near the low-energy $E_1$ and high-energy $E_2$ transitions as well as temperature dependent traces near the $E_1$ transition to show qualitative comparison of decay responses at respective spectral regions and at different temperatures. Each time trace displays a fast decay $\tau_1$ at early times followed by a significantly slower decay $\tau_2$ of the TR response. (c) Fast $\tau_1$ (upper panel) and slow $\tau_2$ (lower panel) decay time constants as a function of probe energy at 10 K and 300 K, respectively. Decay times are extracted from the model fittings described in the main text. Dashed lines are polynomial fits as a guide for the eye, which show the strong variation of decay times with probe energy and temperature.

# Supporting Information

# Revealing Optical Transitions and Carrier Recombination Dynamics within the Bulk Band Structure of Bi$_2$Se$_3$

Giriraj Jnawali,*,† Samuel Linser,† Iraj Abbasian Shojaei,† Seyyedesadaf Pournia,† Howard E. Jackson,† Leigh M. Smith,*,† Ryan F. Need,‡ and Stephen D. Wilson‡

†*Department of Physics, University of Cincinnati, Cincinnati, OH 45221, USA*
‡*Materials Department, University of California, Santa Barbara, CA 93106, USA*

*Corresponding authors:     E-mail: giriraj.jnawali@uc.edu; Ph.: 513-556-0542 (G.J.) &
                            E-mail: leigh.smith@uc.edu; Ph.: 513-556-0501 (L.M.S.)

# Supplementary Note 1: Model fitting parameters

## A. Ground-state parameters:

Table S1: Selected ground-state material parameters of bulk $Bi_2Se_3$.

| Parameters | Values |
| --- | --- |
| Background electron density ($N_0$) | $(7.3 \pm 0.3) \times 10^{18}$ cm$^{-3}$ |
| Background hole density (assumed) | $1 \times 10^{15}$ cm$^{-3}$ |
| Background electron Fermi energy ($E_f$) | 115.5 meV |
| Background hole Fermi energy (assumed) | -11.7 meV |
| Electron mass[1,2] | 0.12 $m_e$ |
| Hole mass[1,2] | 0.24 $m_e$ |
| Refractive index (n)[3] | 5.5 |

The table S1 above lists the selected material parameters for bulk $Bi_2Se_3$ before photoexcitation. These parameters are assumed or taken from the literature, with the exception of the background electron density $N_0$, which is considered a free parameter in the model fitting described in the Methods section. The electron/hole quasi-Fermi energies are measured relative to the conduction band minimum/valence band maximum, with negative values residing in the band gap. The refractive index is approximated as constant within the probe beam energy range.

## B. Fundamental band edge parameters:

Table S2: Fundamental band edge parameters of photoexcited bulk $Bi_2Se_3$.

| Parameters | Values @ 20 ps | Values @ 50 ps |
| --- | --- | --- |
| Electron Fermi energy after photoexcitation ($E_f$) | 117.4 meV | 116.7 meV |
| Photoexcited hole Fermi energy | 3.6 meV | 1.8 meV |
| Photoexcited carrier density ($\Delta N$) | $(1.8 \pm 0.3) \times 10^{17}$ cm$^{-3}$ | $(1.2 \pm 0.3) \times 10^{17}$ |
| Fundamental band gap ($E_{g1}$) | $0.182 \pm 0.005$ eV | $0.182 \pm 0.005$ eV |
| The optical band gap ($E_1$) | 0.358 eV | 0.357 eV |

The Fermi energies are calculated from the total electron/hole density (background plus photoexcited carriers) with an assumed carrier temperature of 30 K (the lattice temperature). The fundamental band gap and photoexcited carrier density are fit as free parameters.

### C. High-energy band edge parameters:

Table S3: High-energy band edge parameters of photoexcited bulk $Bi_2Se_3$.

| Parameters | Values @ 20 ps | Values @ 50 ps |
| --- | --- | --- |
| Electron Fermi energy after photoexcitation ($E_f$) | 117.6 meV | 115.9 meV |
| Hole Fermi energy after photoexcitation | 5.9 meV | 2.3 meV |
| Photoexcited carrier density ($\Delta N$) | $(3.0 \pm 0.8) \times 10^{17}$ cm$^{-3}$ | $(1.4 \pm 0.8) \times 10^{17}$ cm$^{-3}$ |
| High-energy band gap ($E_{g2}$) | $0.985 \pm 0.005$ eV | $0.985 \pm 0.005$ eV |
| High-energy optical gap ($E_2$) | 1.161 eV | 1.159 eV |

As above, the high-energy band gap and photoexcited carrier density are fit as free parameters. The hole mass for $VB_2$ is assumed to be equal to that of $VB_1$, by considering both valence bands are parabolic in the vicinity of gamma point $\Gamma$ (k = 0) of the Brillouin zone.

## Supplementary Note 2: Coherent longitudinal acoustic phonon

Coherent longitudinal acoustic (LA) phonons are propagating ultrasonic strain waves in solid crystals which can be excited efficiently at the surface by ultrafast laser pulses through transient lattice heating and subsequent lattice expansion.[4-6] Since the strain wave propagation modifies the local dielectric properties of the materials, the induced oscillatory behavior of dielectric properties can be probed in the time-domain by various pump-probe spectroscopic techniques.[4-8]

Figure S1a displays probe wavelength dependent transient reflectance signal (TRS) data measured following 1.51 eV pump excitation of the $Bi_2Se_3$ samples and probing at different photon energies (wavelengths). Following initial ultrafast transients, the TRS signal oscillates in time and gradually decays because of interference between light waves reflected off the surface and light waves reflected off the propagating strain wave. The TR signal therefore contains both the electronic response of the material as well as the oscillatory interference signal due to the coherent acoustic modes generated in the sample. The period of the oscillation is nearly 40 ps around the lowest probe wavelength of 1030 nm and it increases up to 60 ps at 1550 nm. The amplitude of the oscillation weakens considerably at longer probe wavelengths and is hardly visible beyond the 1600 nm probe wavelength. Each trace is fitted with bi-exponential functions combined with exponentially decaying cosine function described in the main text and the oscillation period is therefore determined at different probe wavelengths. Figure S1b shows how the period of oscillation changes by changing the probe wavelength. The periods from both 300 K and 10 K data falls in linear regime and the slope $= 1/(2 \cdot n \cdot v_0)$ estimates the average velocity $v_0$ of the LA phonon (sound velocity) with the knowledge of refractive index of the material, $n$. Assuming an average index of refraction of $n = 5$, the sound velocity is calculated to be ~ 2600 m/s, which is reasonable.

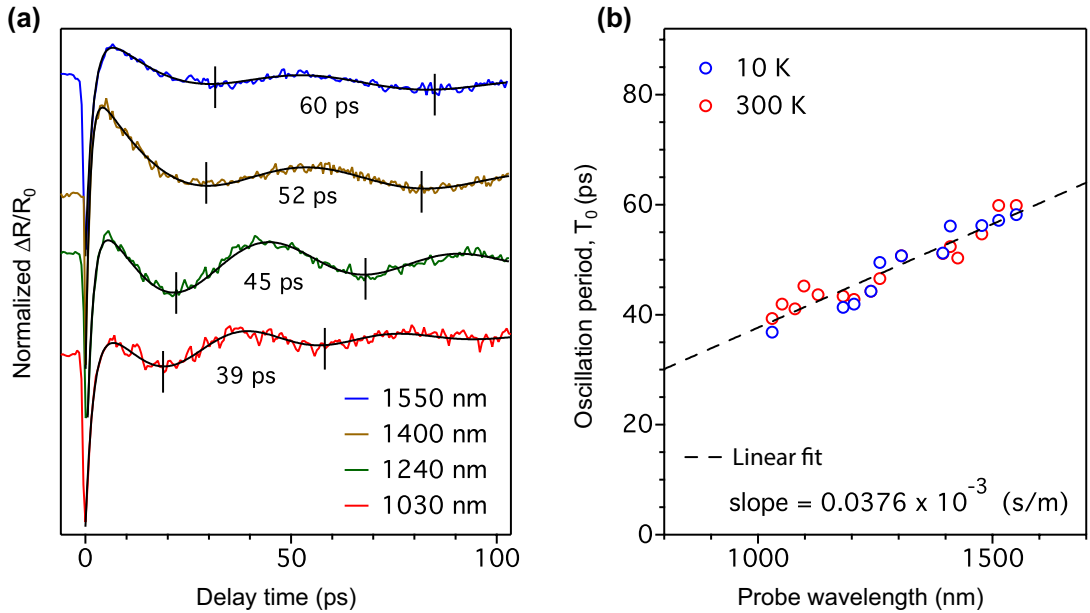

**Figure S1.** Coherent LA phonon as a function of probe wavelength. (a) Normalized time traces of transient reflectance (TR) response of Bi$_2$Se$_3$ flake measured at different probe wavelengths following 820 nm pump excitation at 300 K. Each trace is fitted using bi-exponential functions combined with exponentially decaying cosine function (described in the main text) and oscillation period is estimated at different probe wavelengths. (b) Extracted oscillation periods both at 10 K and 300 K at different probe wavelengths. Dashed black line is a linear fit to the 300 K data which almost coincides with the fit to the 10 K data (not shown), with an estimated slope of 0.037 × 10$^{-3}$ s/m.

## Supplementary Note 3: Estimating band-edge transition at 300 K

It is interesting to note that our room temperature transient reflectance (TR) spectra at the band-edge region does not show substantial changes in terms of broadening and energy when compared to 10 K spectra. This indicates that the simple band-filling model for the absorption does not work at room

S4

temperature. Therefore, we have used a simple model previously applied to fit photoreflectance spectra at excitonic transitions in semiconductors.[9, 10] The model considers that the change in photoreflectance signal is related to the change in the real part of the dielectric function of the sample induced by the pump excitation. The dielectric function at the excitonic transition (near the gap) is considered to have either a Lorentzian or Gaussian profile depending on homogeneous or inhomogeneous broadening. Broadening is caused by the electron-phonon interaction. Therefore, its first derivative reproduces the derivative-like photoreflectance signal. Considering a Lorentzian profile for the dielectric function, the measured normalized TR spectra can be reproduced by using the following expression:[9, 10]

$$\frac{\Delta R}{R} = Re\left[Ae^{i\phi}(E - E_{eff}^0 + i\Gamma)^{-2}\right] \quad (1)$$

where $A$ is the amplitude, $\phi$ is the phase angle, $E_{eff}^0$ is the optical band gap (accounting for band-filling), and $\Gamma$ is the broadening parameter. Terms $A$ and $\phi$ are adjustable constants to fit the data, without any physically relevant meaning. We reproduced each of our TR spectra at 300 K using Equation 5 with reasonable accuracy simply via simple least square fittings (see Figure S2), which allows us to extract the optical band gap of $E_{eff}^0 = 0.36$ eV. If we consider no appreciable change in Fermi energy when cooling down the sample from 300 K and 10 K, the fundamental gap is seen to be nearly the same as the gap deduced at 10 K. We do not consider the energy dependence of the 300 K data to be sufficient to determine the temperature dependence of the band gap.

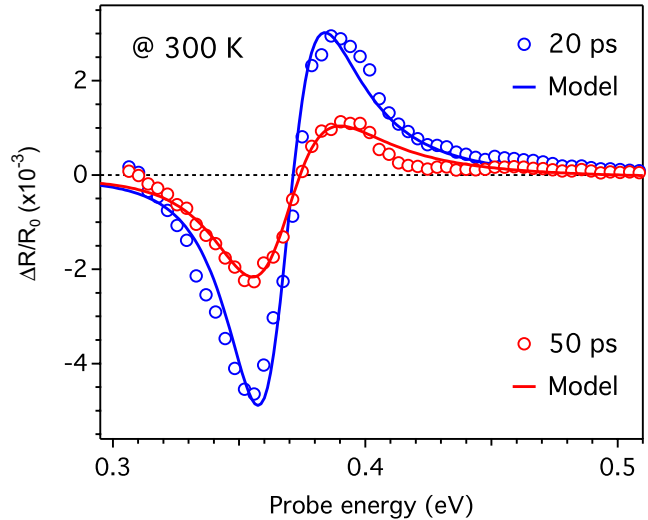

**Figure S2.** Model fitting of TR spectra of a $Bi_2Se_3$ nanoflake at 300 K. Selected spectra at 20 ps and 50 ps delay times are reproduced by using a model (Eq. 1). Least square fitting calculates effective band gap (optical gap) of ~ 0.36 eV at 300 K, which is nearly identical with the optical gap measured at 10 K.

S5

# Supplementary Note 4: *I-V* characteristics and photocurrent spectroscopy

$Bi_2Se_3$ flakes are exfoliated from a single crystal using the scotch tape method and dispersed onto p+ doped silicon substrates which have a 300 nm insulating oxide layer. Relatively large and thicker flakes are chosen for photocurrent measurements. Morphology and thickness of the flake are characterized after the measurements by using atomic force microscope in the tapping mode. Typical thicknesses range from 200 – 500 nm, which are comparable with the thickness of the flakes used in the transient reflectivity measurements.

Large metallic contact pads of 300 nm Aluminium (Al) on 20 nm Titanium (Ti) are fabricated on either side of the flakes using a standard photolithographic process. Before metal deposition, the sample was immersed in a solution of HF (1) : $H_2O$ (7) for few seconds to remove oxides and photoresist contaminants, which allow to achieve an ohmic contact between the flake and the metal contacts. The sample is mounted onto the cold finger of a $He^4$ constant-flow cryostat (Janis ST-500), which allows cooling of the sample down in vacuum to 10 K. Before photocurrent (PC) measurements, all devices are checked by acquiring *I-V* curves in dark both at 300 K and 10 K. A linear *I-V* characteristic as shown in Figure S3 a indicates each of our devices is metallic in nature.

Since the devices are highly conducting, we used a small constant current $I$ =1 nA to measure a weak PC signal. The PC signal is collected as a small light-induced modulation of the voltage bias (amplified by 100 times) using a lock-in amplifier with the chopper synchronized signal as a reference. The laser beam is focused using a 50× reflective objective perpendicularly onto the device (focus diameter = 2 µm) during PC measurements. The PC spectra are obtained both at RT and LT (10 K) by collecting the PC signal at different wavelengths in the mid-IR range (3000 – 4000 nm) of the incident laser pulse. The PC spectra are shown in Figure S3 b. Each spectrum is normalized by the number of incident photons. An important feature is the rapid increase of PC signal after around 0.3 eV, which is nearly temperature independent. The increase of PC signal is attributed to increase of free carriers near Fermi level via band-edge optical absorption in $Bi_2Se_3$. The maxima are reached at around 0.36 eV, which can be considered to be the optical band gap of $Bi_2Se_3$.

Despite the simple device structure, the main features of the measurements, particularly the linear *I-V* curves, are consistent with a highly doped $Bi_2Se_3$ sample. Moreover, a temperature independent optical gap of 0.36 qualitatively supports TRS results. The observation of a PC signal from a highly metallic $Bi_2Se_3$ strongly suggests the bulk response is indeed dominant in our samples.

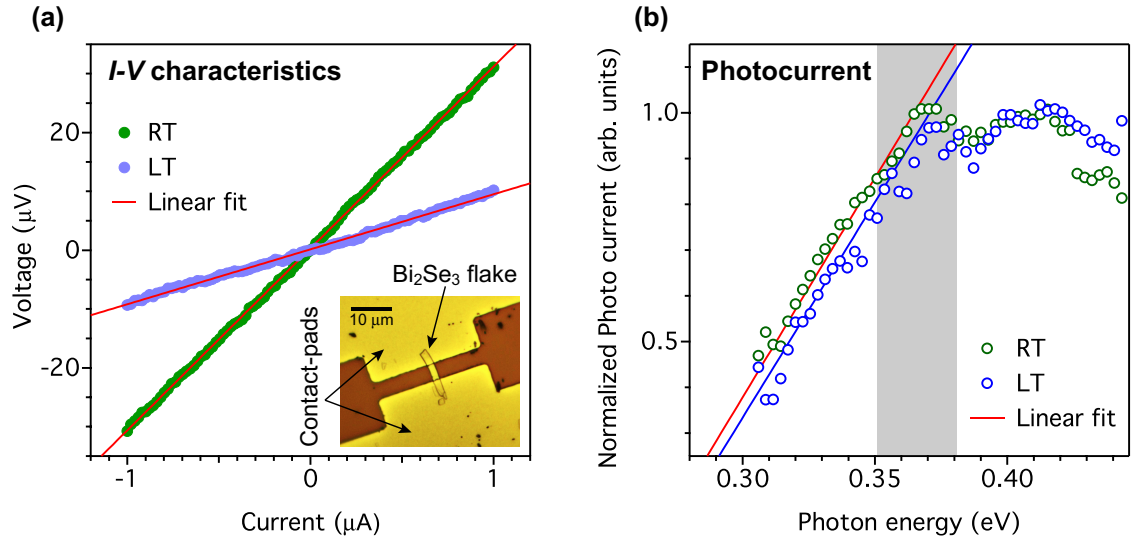

**Figure S3.** Mid-IR photoconductive response of a single $Bi_2Se_3$ nanoflake device. (a) Dark *I-V* characteristics, along with linear fits (red lines), of a single $Bi_2Se_3$ nanoflake device at RT (light green sphere) and LT (light blue sphere), respectively. The linear behavior in *I-V* curves indicates metallic nature the $Bi_2Se_3$ device. Inset shows an optical image of the flake with two Ti/Al contacts at either end. (b) Photocurrent (PC) of the device at RT (green circles) and low temperature (LT ~ 10 K) (blue circles) as a function of laser excitation photon energy at a constant DC current of 1 nA. PC signal is normalized to number of incident photons on the device and plotted by further normalizing the data with maximum of PC signal. Shaded rectangle shows estimated optical band gap for the device, which is consistent with the transient reflectivity measurements described in the main text.